\documentclass[journal,12pt,draftclsnofoot,onecolumn]{IEEEtran}
\usepackage[dvips]{graphicx}
\usepackage{amssymb}
\newcommand{\bA}{\mathbf{A}}
\newcommand{\bx}{\mathbf{x}}
\newcommand{\by}{\mathbf{y}}
\newcommand{\br}{\mathbf{r}}

\begin{document}
\title{Nonlinear self-flipping of polarization states in asymmetric waveguides}

\author{
Wen Qi Zhang, M. A. Lohe, Tanya M. Monro and Shahraam Afshar V.
\thanks{%Manuscript received 
%April 16, 2014; revised January 11, 2010;
%accepted July 54, 2020. Date of publication July 33, 2032.
This work was supported in part by the ARC (project DP110104247).
T.~M.~Monro acknowledges the support of an ARC Federation Fellowship.
}%
\thanks{The authors are with the Institute for Photonics \& Advanced Sensing 
(IPAS), and the Department of Physics,
The University of Adelaide, 5005, Australia}%
%\thanks{Color versions of the figures in this letter 
%are available online at http://ieeexplore.ieee.org.}%
%\thanks{Digital Object Identifier 10.1109/LPT.2011.21788??}%
}

%\markboth{IEEE PHOTONICS TECHNOLOGY LETTERS,~Vol.~6?, No.~1?, January~2099}%
%{Shell \MakeLowercase{\textit{et al.}}: Bare Demo of IEEEtran.cls for Journals}

\maketitle

\begin{abstract}
Waveguides of subwavelength dimensions with asymmetric geometries, such as 
rib waveguides, can display nonlinear polarization effects in which
the nonlinear phase difference dominates the linear contribution, 
provided the birefringence is sufficiently small. We demonstrate that 
self-flipping polarization states can appear in such rib waveguides at low (mW) 
power levels. We describe an optical power limiting device with
optimized rib waveguide parameters that can operate at low powers with
switching properties.
\end{abstract}

%\begin{IEEEkeywords}
%Rib waveguides, nonlinear optical waveguides, optical waveguide polarization, 
%nonlinear pulse propagation, nonlinear optical devices.
%\end{IEEEkeywords}

\IEEEpeerreviewmaketitle

\section{Introduction}

\IEEEPARstart{N}{onlinear}
interactions between the two polarization modes of a waveguide 
lead to intriguing physical effects and
opportunities for new devices for optical data processing 
%\cite{Wabnitz09,Kozlov11,Fatome10,Agrawal07}.
\cite{Wabnitz09}-\cite{Agrawal07}.
We show here how nonlinear interactions of this form can be employed to develop 
all-optical devices at low power levels, using
switching (bistable) properties of the polarization phase difference,
as described in \cite{Zhang11}, 
at zero birefringence wavelength in subwavelength rib waveguides. 
In general, the interaction between the two
polarizations in an optical waveguide is described by the 
coupled Schr\"odinger equations 
\cite{Agrawal07}:
\begin{eqnarray}
&&\frac{\partial A_{j}}{\partial z} +\sum_{n=1}^{\infty}\frac{i^{n-1}}%
{n!}\beta_{jn}\frac{\partial^{n}A_{j}}{\partial t^{n}}
\label{eq01}
\\
=&& \hspace{-6mm}
i\,\left(\gamma_{j}\left\vert A_{j}\right\vert ^{2}+\gamma_{c} \left\vert
A_{k}\right\vert ^{2}\right)A_{j}+i\gamma_{c}^{\prime}A_{j}^{\ast}A_{k}^{2}%
\exp(-2iz\Delta\beta_{jk}),
\nonumber
\end{eqnarray}
where $j,k=1,2\;(j\ne k)$ index the two polarization modes, $A_{1},A_{2}$ are
the amplitudes of the corresponding fields, $\beta_{jn}$ are the $n$th order
propagation constants, $\Delta\beta_{jk}=-\Delta\beta_{kj}$ is the linear
birefringence, $\gamma_{j},\gamma_{c}$ and $\gamma_{c}^{\prime}$ are the
effective nonlinear coefficients representing self phase modulation, cross
phase modulation and coherent coupling of the two polarization modes,
respectively. 

The weak guidance approximation assumes that the effect\-ive mode
areas of the two polarization modes are equal \cite{Agrawal07}, leading to
$\gamma_{1} = \gamma_{2} = 3\gamma_{c}/2 = 3\gamma_{c}^{\prime}$.
We have shown in \cite{Zhang11,SLZM} 
that these equalities are not necessarily valid
for waveguides with large index contrast and 
subwave\-length dimensions. For such cases a
new class of time-independent polarization states appear, 
in which the phase difference between the two
polarizations oscillates abruptly between two well-defined values,
which we refer to as switching (bistable) behavior \cite{Zhang11}.
Associated with this, the polarization state 
of the propagating mode flips abruptly through fixed angles. 
This is the result of competition between the linear and
nonlinear phase differences of the two polarization modes propagating
along the waveguide. While the linear phase difference is
proportional to $z\Delta\beta$ (where $\beta=\beta_{12}=-\beta_{21}$), 
the nonlinear phase difference at
any point along the waveguide depends on the $\gamma$ values and on the 
power coupled into each polarization. As a result, one expects that
significant nonlinear polarization effects should be observed as
$\Delta\beta$ approaches zero, i.e.\ when the linear phase difference is
negligible. In waveguides with highly symmetric geometries, however, we
have $\Delta\beta=0$ only if there is also a symmetry
between the polarization mode distributions
(for example, in waveguides with circular or square cross sections), 
in which case the nonlinear phase difference also approaches zero for 
the same power coupled into the two polarizations. For this reason, we
previously found polarization switching only for 
powers at the kW level \cite{Zhang11}.

%%%%%%%%%%%%%%%%%%%%%%%%%%%%%%%%%%%%%%%%%%%%%%%%%%%%%%%%%%%%%%%%%%%%%%%%%%%%%%%%
%%%%%%%%%%%%%%%%%%%%%%%%%%%%%%%%%%%%%%%%%%%%%%%%%%%%%%%%%%%%%%%%%%%%%%%%%%%%%%%%

\section{Rib waveguides}

Here, we consider rib waveguides with structural
dimensions chosen such that the birefringence is nearly zero at the operating 
wavelength of 1550nm, following 
%\cite{VLD,HRH,Chan05}. 
\cite{VLD}-\cite{Chan05}. 
For such cases, the waveguides are polarization independent i.e., 
$\Delta\beta=0$. Previous studies have identified the relation 
between the width and the height (etch depth) of the rib for 
independent polarization operation of rib waveguides \cite{VLD,Chan05} and 
have used them in applications such as ring resonators \cite{HRH}. 
For these waveguides, the mode field 
distributions for the two polarizations are not identical, which leads to 
$\gamma_{1}\ne\gamma_{2}$.  Therefore, the nonlinear phase 
difference is large relative to the linear contribution, and 
leads to switching behavior at power levels several 
orders of magnitude lower than before \cite{Zhang11}. 
Hence, we are able to demonstrate polarization self-flipping through 
$\pi/2$ at just mW powers. This approach produces polarization 
states which flip under their own power, whereas
in \cite{Kozlov11,Fatome10} a second pump is required. 

%%%%%%%%%%%%%%%%%%%%%%%%%%%%%%%%%%%%%%%%%%%%%%%%%%%%%%%%%%%%%%%%%%%%%%%%%%%%%%%%
\begin{figure}[t]
\centering
\includegraphics[width=\columnwidth]{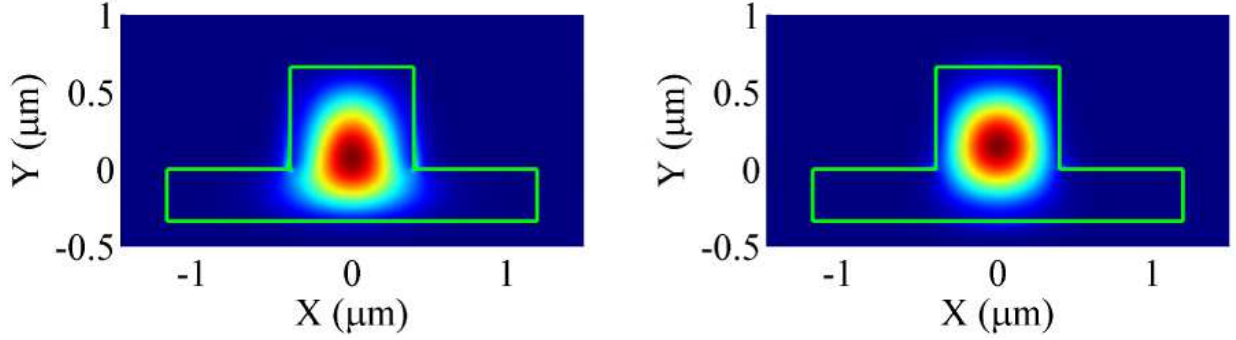}
\caption{
Mode profiles for a rib waveguide with dimensions as shown.
}
\label{fig1}
\end{figure}
\begin{figure}[!hb]
\centering
\includegraphics[width=\columnwidth]{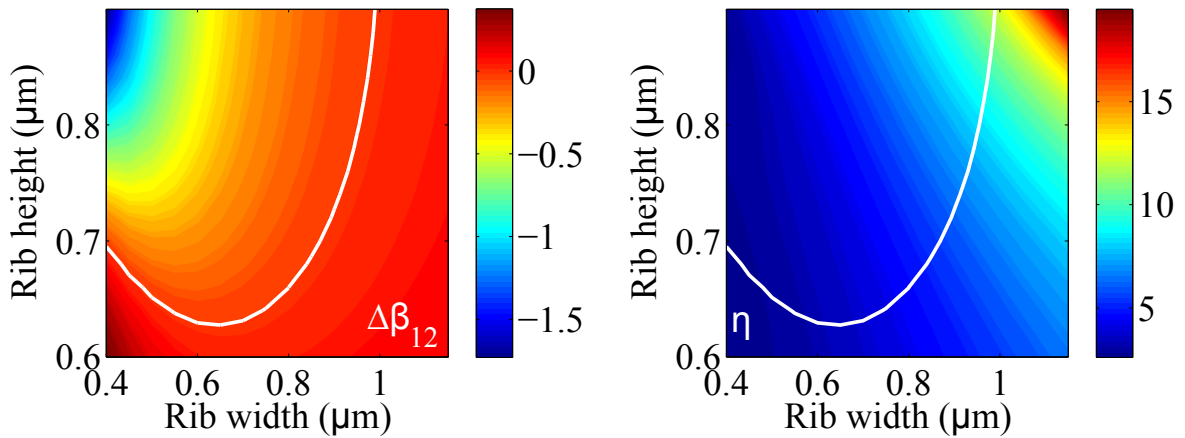}
\caption{
Contour plot of $\Delta\beta$ in m$^{-1}$ (left), 
and the efficiency factor $\eta$ (right) defined in Eq.\ (\ref{e7}),
for a range of rib waveguide parameters.
}
\label{fig2}
\end{figure}

%%%%%%%%%%%%%%%%%%%%%%%%%%%%%%%%%%%%%%%%%%%%%%%%%%%%%%%%%%%%%%%%%%%%%%%%%%%%%%%%
An example of mode field distributions for two 
polarizations with $\Delta\beta=0$
is shown in Fig.\ \ref{fig1}, for a chalcogenide glass-based rib 
waveguide.
Evidently the distributions for the two polarizations are not identical,
which leads to $\gamma_{1}\ne\gamma_{2}$.
These profiles have been calculated using a finite element package,
and the $\gamma$ values have been determined using the formulas (3) in
\cite{Zhang11}. The birefringence can be calculated
as a function of the rib width and rib height (i.e.\ etch depth), 
and is plotted in Fig.\ \ref{fig2},
where the white line indicates points for which $\Delta\beta=0$. 
%%%%%%%%%%%%%%%%%%%%%%%%%%%%%%%%%%%%%%%%%%%%%%%%%%%%%%%%%%%%%%%%%%%%%%%%%%%%%%%%
%%%%%%%%%%%%%%%%%%%%%%%%%%%%%%%%%%%%%%%%%%%%%%%%%%%%%%%%%%%%%%%%%%%%%%%%%%%%%%%%

We now investigate nonlinear polarization as
described in \cite{Zhang11,SLZM}, for waveguides with $\Delta\beta\approx0$
(near the white line in Fig.\ \ref{fig2}, left). 
We substitute $A_{j}=\sqrt{P_{j}}\,e^{i\phi_{j}}$ 
for $j=1,2$ into Eqs.~(\ref{eq01}), where $P_{j}$ is the power of the
field $A_{j}$ with phase $\phi_{j}$ and look for time-independent solutions. 
The total power $P_0=P_{1}+P_{2}$ is constant in $z$. 
We define the following dimensionless variables:
\begin{eqnarray}
v =\frac{P_{1}}{P_{0}}, \qquad\theta=2\Delta\phi, &&
\tau=2\gamma_c^{\prime}P_0z,
\label{g0}
\\
a=-\frac{\Delta\beta}{\gamma_{c}^{\prime}P_{0}}-%
\frac{\gamma_{c}-\gamma_{2}}{\gamma_{c}^{\prime}}, 
&& b=\frac{\gamma_{1}+\gamma_{2}-2\gamma_{c}}{2\gamma_{c}^{\prime}},
\label{g1}%
\end{eqnarray}
where $\Delta\phi=\phi_{1}-\phi_{2}+z\Delta\beta$ is the phase difference
between the two fields. 
From Eqs.~(\ref{eq01}) we obtain
a system of equations for $v,\theta$ (see Eqs.\ (5,6) in \cite{Zhang11}) 
which can be solved for any given $v_0=v(0),\theta_0=\theta(0)$ 
at the initial location $\tau=0$ (with $0<v_0<1$),
either analytically, or numerically for any
specified values of $a,b$. There are three types of solutions:
steady state
solutions, periodic (including switching) solutions, and soliton solutions. 
For periodic solutions the dimensionless period $T_{\tau}$ of $v,\cos\theta$
(regarded as functions of $\tau$) depends only on $a,b$ and
on $v_0,\theta_0$. Hence, the functions $P_1,P_2,\cos2\Delta\phi$ are
periodic in $z$ with a period $T_z=T_{\tau}/(2\gamma_c'P_0)$, as follows
from the definition of $\tau$ in Eq.\ (\ref{g0}). 
The switching behavior discussed in \cite{Zhang11,SLZM} requires
$1<a<2b-1$ which from the definitions (\ref{g1}) implies that
\begin{equation}
\label{e6}
P_0(\gamma_c+\gamma_c'-\gamma_1)
<\Delta\beta< P_0(\gamma_2-\gamma_c-\gamma_c').
\end{equation}
This inequality is satisfied at relatively low powers $P_0$ provided
$\Delta\beta$ is sufficiently small.

Let $L$ be the length of the waveguide, hence $0\leqslant z\leqslant L$.
We are interested in exploring the possible values of $v,\theta$ 
at the endpoint $z=L$, 
i.e.\ at $\tau=2\gamma_c^{\prime}P_0L$; $v,\theta$ each depend 
nontrivially on the total power $P_0$, regarded as a variable parameter, 
through $a$. 
One can in principle determine the explicit dependence of $v,\theta$ on
$P_0$ with the help of the exact solutions, but
for waveguides with $\Delta\beta=0$, however, both parameters $a,b$ are 
fixed and in this case $P_0$
enters the defining dim\-ensioned equations as a scale factor only.
This means that the periodic functions $v,\cos\theta$ and their 
period $T_{\tau}$ are determined entirely by the 
values of the $\gamma$ coefficients, together with the initial values 
$v_0,\theta_0$.  The functions $P_1(z),P_2(z),\cos\Delta\phi(z)$ at $z=L$
can therefore be stretched or compressed by varying either $P_0$ or $L$, 
in particular $\cos\Delta\phi$ at $z=L$ may be regarded as a function of 
either $L$ or $P_0$ and, since $2\Delta\phi=\theta(2\gamma_c'P_0L)$,
is periodic in both $L$ and $P_0$. The corresponding periods $T_L$ and
$T_{P_0}$ of $\cos2\Delta\phi$, of dimension length and power respectively, 
are given by 
\begin{equation}
\label{eq04}
T_L=T_z=\frac{T_{\tau}}{2\gamma_c'P_0},
\qquad
T_{P_0}=\frac{T_{\tau}}{2\gamma_c'L},
\end{equation}
where the dimensionless period $T_{\tau}$ is independent of both $L,P_0$. 
These simple dependencies, a consequence of the scaling properties, 
are useful for the application outlined below.

We consider now light that is initially linearly polarized, i.e.\ 
$\theta_0=0,$ with equal power coupled into the two polarizations, 
hence $v_0=1/2$. 
As an example, as explained further below, we choose
a rib width of $990$ nm and a rib height (i.e.\ etch depth) of $900$ nm, 
with a substrate width of $2400$ nm and
a height of $340$ nm. We find, in units of (W.m)$^{-1}$:
$\gamma_1=81.6, \gamma_2=82.2, \gamma_c=49.6,\gamma_c^{\prime}=24.7$, with
numerical values for $\Delta\beta$ such that
$|\Delta\beta|\leqslant 3.8\times 10^{-4}$ m$^{-1}$.
With $1$ mW input power and a $1$ m long waveguide, 
the nonlinear phase change is more than 200 times larger than
the linear contribution. 

\begin{figure}[!ht]
\centering
\includegraphics[width=\columnwidth]{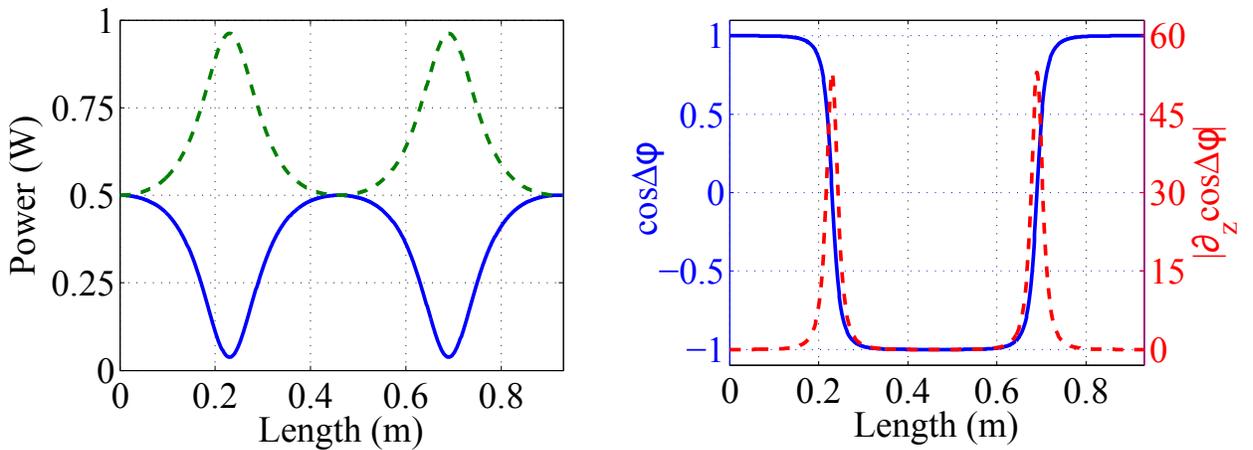}
\caption{
$P_1$ (blue), $P_2$ (green) on the left, and
$\cos\Delta\phi$ (blue), $|\partial_z\cos\Delta\phi|$ (red) 
as functions of $z$.  
}
\label{fig3}
\end{figure}

With these $\gamma$ values we have $a=1.32,b=1.308$, and so
$1<a<2b-1$ is satisfied, and therefore switching states exist. 
Fig.\ \ref{fig3} shows 
$P_1,P_2,\cos\Delta\phi$ as functions of $z$ at $P_0=1$ W.
Evidently $P_1+P_2$ is constant, and
$P_1,P_2$ each have the period $T_z=0.47$ m;
$\cos\Delta\phi$ displays switching by flipping between $\pm1$
with a period $2T_z=0.94$ m. 
%%%%%%%%%%%%%%%%%%%%%%%%%%%%%%%%%%%%%%%%%%%%%%%%%%%%%%%%%%%%%%%%%%%%%%%%%%%%%%%%
%%%%%%%%%%%%%%%%%%%%%%%%%%%%%%%%%%%%%%%%%%%%%%%%%%%%%%%%%%%%%%%%%%%%%%%%%%%%%%%%

We optimize the rib waveguide parameters by choosing 
values that result not only in $\Delta\beta=0$, 
but also such that $\cos\Delta\phi$ 
flips as abruptly as possible between $\pm1$, relative to the period.
In order to quantify this transition, 
we define a transition length $L_{\mathrm{trans}}$
over which $\cos\Delta\phi$ flips between $\pm1$,
and a dimensionless efficiency factor 
$\eta$ by
\begin{equation}
L_{\mathrm{trans}}=\frac{2}{\left\vert \partial_{z}
\cos\Delta\phi(z)\right\vert
_{\cos\Delta\phi=0}}, 
\qquad
\eta=\frac{T_z}{L_{\mathrm{trans}}},
\label{e7}%
\end{equation}
where $2T_z$ is the period of $\cos\Delta\phi(z)$. 
The derivative 
$\partial_{z}\cos\Delta\phi(z)$ is evaluated at $\cos\Delta\phi=0$ 
since this is where
the maximum slope of $\cos\Delta\phi$ occurs. This is evident in
Fig.\ \ref{fig3} (right), which plots $|\partial_{z}\cos\Delta\phi(z)|$ 
(red) as a function of $z$. 

For low power operations and fast polarization
transitions we minimize $L_{\mathrm{trans}}$ relative
to the period $2T_z$, i.e.\ we  maximize $\eta$ within the physical
limit of a unit length waveguide, by varying the rib width and height but
keeping $\Delta\beta=0$. This is equivalent to minimizing
$|a-b|$, which ensures that the initial value $v_0=1/2$ lies
close to the unstable steady state at $v=(a-1)/2(b-1)$, as in the example
shown in Figs.\ 5(ii) and 6(ii) of \cite{SLZM}.
Fig.\ \ref{fig2} (right) shows $\eta$ for a range of rib waveguides
of varying rib width and rib height, where $\Delta\beta=0$ holds
on the white line. 
The maximum value of $\eta$ occurs at the top right of the white line, with
larger values
corresponding to periods $2T_z$ that exceed the length $L$
of the waveguide. Choosing $L\approx1$ m, we arrive at the rib width 
$990$ nm and rib height $900$ nm as considered above, with $\eta=12.3$.

%%%%%%%%%%%%%%%%%%%%%%%%%%%%%%%%%%%%%%%%%%%%%%%%%%%%%%%%%%%%%%%%%%%%%%%%%%%%%%%%
%%%%%%%%%%%%%%%%%%%%%%%%%%%%%%%%%%%%%%%%%%%%%%%%%%%%%%%%%%%%%%%%%%%%%%%%%%%%%%%%

\section{Power limiting optical device}
The switching properties of $\cos\Delta\phi$, considered
as a function of either $P_0$ or the waveguide length $L$, enable us to 
construct in principal a power limiting optical device (a ``surge protector").
Consider a rib waveguide with the structural parameters of
the given example, with a polarizer at the output (at $z=L$) 
which is aligned either
parallel or perpendicular to the direction of polarization of the input beam,
which is linearly polarized. Denote the vector field at $z=L$ by
$\bA_{\mathrm{out}}=A_1\widehat{\mathbf{x}}+A_2\widehat{\by}$, where 
$\widehat{\bx},\widehat{\by}$ are the unit vectors along the $X,Y$ axes
respectively, and let $\widehat{\br}=(\widehat{\bx}+\widehat{\by})/\sqrt{2}$ be
the unit vector defining the direction of the polarization axis of the 
polarizer.
The polarizer therefore projects the output vector field onto 
$\bA_{\mathrm{out}}\centerdot\widehat{\br}
=\sqrt{P_1/2}\,e^{i\phi_1}+\sqrt{P_2/2}\,e^{i\phi_2}$, with a 
power output 
$P_{\mathrm{out}} = |\bA_{\mathrm{out}}\centerdot\widehat{\br}|^2$.
As a function of the initial total power $P_0$ we have
$P_{\mathrm{out}}(P_0)=P_0/2+P_0\sqrt{v(1-v)}\cos(\theta/2)$ where
$v,\theta$ are known functions of $\tau=2\gamma_c^{\prime}P_0L$.

\begin{figure}[hbt]
\centering 
\includegraphics[width=3.2in]{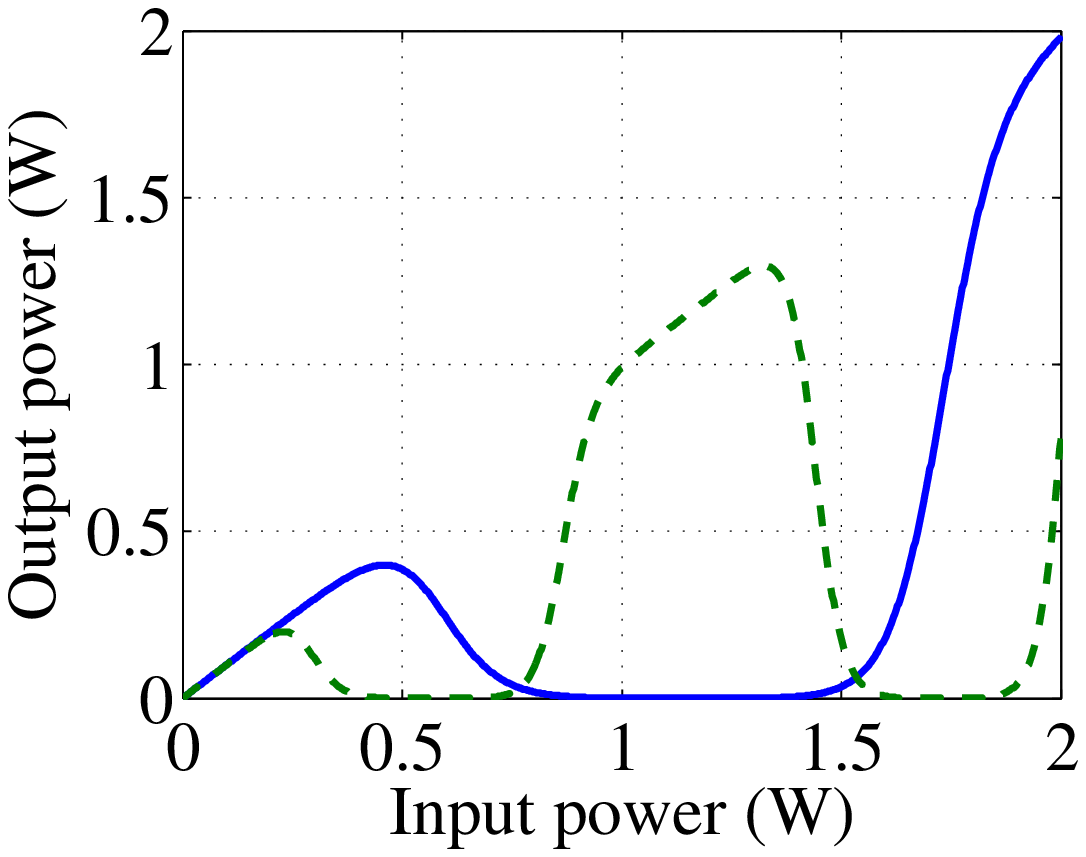}%
\includegraphics[width=3.2in]{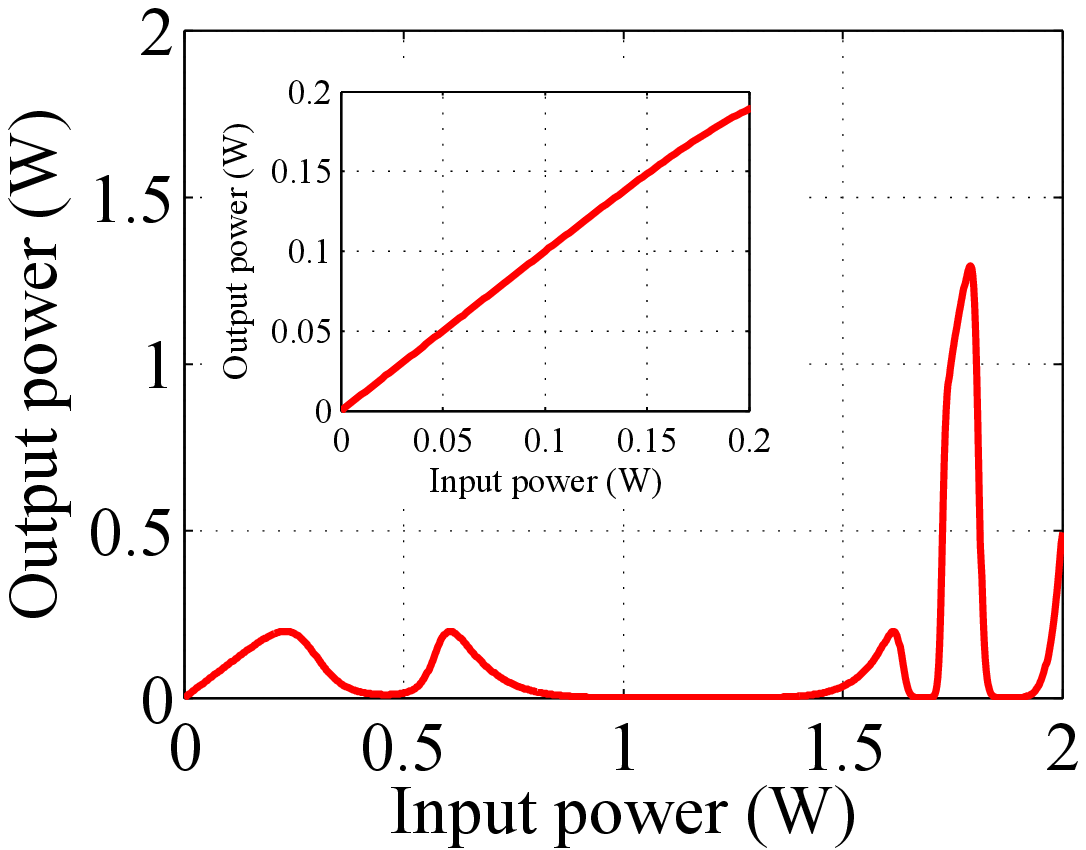}
\caption{
The power output $P_{\mathrm{out}}$
for waveguide lengths $0.4$ m (blue) and $0.8$ m (green), each
with polarizers at the output end,  
and the combined two-stage output (red).
}
\label{fig4}
\end{figure}

We choose $L=L_1=0.4$ m, so that switching behavior appears at
power levels $P_0\approx1$ W. The output power
$P^1_{\mathrm{out}}$ is shown in Fig.\ \ref{fig4} (blue) 
as a function of $P_0$, and depends linearly
on $P_0$ for $P_0\leqslant0.3$ W, but falls to zero in the range $0.9-1.4$ W.
Hence the polarizer in conjunction with the switching properties 
reduces the power output to zero over this range, and yet has no effect
for small powers, i.e.\ $P^1_{\mathrm{out}}(P_0)=P_0$ for $P_0\leqslant0.3$ W.
%%%%%%%%%%%%%%%%%%%%%%%%%%%%%%%%%%%%%%%%%%%%%%%%%%%%%%%%%%%%%%%%%%%%%%%%%%%%%%%%
%%%%%%%%%%%%%%%%%%%%%%%%%%%%%%%%%%%%%%%%%%%%%%%%%%%%%%%%%%%%%%%%%%%%%%%%%%%%%%%%

In order to extend the range over which
the output power is limited, we consider a second waveguide with the same 
structural parameters as before, but of
length $L_2=2L_1=0.8$ m, with an output $P^2_{\mathrm{out}}$ as shown
in Fig.\ \ref{fig4} (green); this plot is 
a compressed copy of the plot for $P^1_{\mathrm{out}}$, as follows from the
scaling properties with respect to $P_0$ and $L$.  This
second waveguide and polarizer are attached to the output
end of the first waveguide,
and the first pol\-arizer then ensures that the initial conditions for the 
second waveguide
are $\theta_0=0,v_0=1/2$. The total output of the device as a function of $P_0$ 
is given by $P^2_{\mathrm{out}}(P^1_{\mathrm{out}}(P_0))$, which is shown in
Fig.\ \ref{fig4} (red). The output is linear for
$P_0\leqslant0.15$ W, but is limited to values $\leqslant0.2$ W for all input
powers $P_0\leqslant1.7$ W.
The maximum power operation range can be adjusted
by varying the lengths and hence periods of the connected waveguides,
and also by adding further stages.
This conceptual design does not require any additional pumps for its
operation, which is characteristic of the structurally-induced 
nonlinear pol\-arization effects discussed previously \cite{Zhang11,SLZM}. 

%%%%%%%%%%%%%%%%%%%%%%%%%%%%%%%%%%%%%%%%%%%%%%%%%%%%%%%%%%%%%%%%%%%%%%%%%%%%%%%%
%%%%%%%%%%%%%%%%%%%%%%%%%%%%%%%%%%%%%%%%%%%%%%%%%%%%%%%%%%%%%%%%%%%%%%%%%%%%%%%%

\section{Conclusion}
We have demonstrated the existence of nonlinear phase effects 
for waveguides with reduced symmetries, which are much larger
than linear effects, at mW power levels.
We have also demonstrated that switching behavior with $\pi/2$ polarization 
flipping can occur at the $1$ W input power level. 
Although we have analyzed properties of the system for the case of
zero birefringence, this 
is convenient but not essential; the exact solutions described in 
\cite{Zhang11,SLZM} allow one to determine the precise dependence of 
$\cos\Delta\phi$ on the input power for any $\Delta\beta$, including
small nonzero values. The vectorial nonlinear model for subwavelength
waveguides has enabled us to 
uncover these phenomena. Our results show that
applications displaying this nonlinear polarization behavior can, in principle,
operate at practical power levels.

%\section*{Acknowledgment}

%T.~M.~Monro acknowledges the support of an ARC Federation Fellowship.

\ifCLASSOPTIONcaptionsoff
  \newpage
\fi



%\bibliographystyle{IEEEtran}

\end{document}